\documentstyle[12pt]{article}
\def\smal{}
\def\sss{}
\def\Comm#1{{\tt{[#1]}}}

\def\R{\rm R}
\def\th{\theta}
\def\ep{\epsilon}
\def\szi{{\sum_{m=0}^\infty}}

\def\szit{{\sum_{l,m=0}^\infty}}

\def\sonwm{{\sum_{m=1}^{n_1-1}}}
\def\sontm{{\sum_{m=1}^{n_2-1}}}
\def\sonam{{\sum_{m=1}^{n_a-1}}}
\def\pzmm{\prod_{l=1}^{m-1}}
\def\X{{\bf x}}
\def\a{{\bf w}}
\def\M{{\it M}}
\def\pl{\partial}
\def\ba{{\bf y}}
\def\om{\omega}
\def\OM{W}
\def\q{\quad}

\def\ie{{\it i.e.}}

\catcode`@=11
\newdimen\z@ \z@=0pt
\def\m@th{\mathsurround=\z@}
\def\ialign{\everycr{}\tabskip\z@skip\halign} 
\def\eqalign#1{\null\,\vcenter{\openup\jot\m@th
  \ialign{\strut\hfil$\displaystyle{##}$&$\displaystyle{{}##}$\hfil
      \crcr#1\crcr}}\,}
\def\matrix#1{\null\,\vcenter{\normalbaselines\m@th
    \ialign{\hfil$##$\hfil&&\quad\hfil$##$\hfil\crcr
      \mathstrut\crcr\noalign{\kern-\baselineskip}
      #1\crcr\mathstrut\crcr\noalign{\kern-\baselineskip}}}\,}
\catcode`@=12
\def\be{\begin{equation}}
\def\ee{\end{equation}}
\def\eq#1{{(\ref{#1})}}

\begin{document}
\textheight 9in
\topmargin 0in
\begin{flushright}
DAMTP/97-13
\end{flushright} 
\vskip20pt
\begin{center}
\begin{Large}
{\bf A braided interpretation of fractional }\vskip0pt 
{\bf supersymmetry in higher dimensions}
\\
\end{Large}
\vspace{1cm}
{\bf R.S. Dunne\footnote{e-mail: r.s.dunne@damtp.cam.ac.uk}}
\vskip10pt
{\it Department of Applied Mathematics \& Theoretical Physics}\vskip0pt
{\it University of Cambridge, Cambridge CB3 9EW}
\\
\end{center}
\begin{abstract}
A many variable $q$-calculus is introduced using the formalism of
braided covector algebras. Its properties when certain of its
deformation parameters are roots of unity are discussed in detail,
and related to fractional supersymmetry. The special cases of two 
 dimensional supersymmetry and fractional supersymmetry are 
developed in detail.\vskip20pt\noindent

\end{abstract}
\vfill\eject
\section{Introduction} In four recent papers
\cite{DMAP1,DMAP2,DMAP3,DMAP4} the properties of the braided line \cite{MajI,MajII} when
its deformation parameter is a root of unity were discussed. Most
notably these studies led to a novel understanding of one dimensional
supersymmetry and fractional supersymmetry \cite{AM,ABL,Kerner}. Our aim in
the present paper is to extend these results to the many variable case.
In order to do this we  
construct a many variable, generic $q$, generalised Grassmann algebra using the formalism of
braided covector algebras \cite{MajI,MajII}  with suitable $\R$ and
$\R'$ matrices. Within the framework provided by this
formalism the construction of the corresponding many variable left $q$-calculus is straightforward.
After a little further work the corresponding right $q$-calculus is
also obtained.
\par
In this many variable case it is convenient to generalise the graded
brackets used in \cite{DMAP2}
to a pair of braided brackets (left and right), which we introduce in section 2.
This change has several, in general useful, consequences. In particular, left and right
differentiation and integration become truly distinct, rather than being the same thing induced by
different algebraic operators as was the case in \cite{DMAP2}. There are well defined and simple 
commutation relations between all of these operations. Another advantage over the approach of
\cite{DMAP2} is that the conditions which govern the commutation relations
of non-commuting constants are built into the many variable algebra, so that they are no longer
additional constraints. In contrast to the situation with graded
brackets, the conditions necessary in order that left and right
differentiation be induced are compatible. 
A consequence of this is that we can now work quite generally with both left and right
differentiation/integration, rather than being restricted to one or
the other, as was the case 
for graded bracket based $q$-calculus.
\par 
In section 3 we take the $q_a\to{\tilde q}_a$ limit  (${\tilde q}_a$ a root of unity) of the many
variable $q$-calculus, and obtain many variable analogues of the structures and decompositions 
seen in \cite{DMAP2,DMAP3,DMAP4}. A full set of commutation relations between the different algebraic
elements, derivatives and integrals is also given. At the end of this section the braided Hopf
structure of both the coordinates and the derivatives is given, as well as the duality between
them. Further details of this duality, as well as an alternative discussion of the braided line
Hopf algebra (at generic $q$ and at $q$ a root of unity) are give in the appendix.
\par
 Section 4 deals with the case of two dimensional supersymmetry. This plays an important
role in superstring theory \cite{GSW}, in which it corresponds to supersymmetry on the world sheet of the
string. The full two dimensional supersymmetry algebra and transformations are recovered, and all
of the transformation properties of the bosonic spacetime variables $x$ and $t$ emerge as
consequences of their definition as different combinations of the $q_a\to-1$ limits of two braided
line coordinates $\{\theta_a\}$ (a=1,2). Together these two braided lines make up a braided plane,
but we note that this is not the braided/quantum plane which is usually encountered in the
literature \cite{MajI,MajII}. In the limit, translations within this braided plane induce supersymmetry
transformations of $x$ and $t$. Furthermore, once the Lorentz transformations of $\theta_1$ and
$\theta_2$ are specified, those of $x$ and $t$ follow automatically. The results are in agreement
with the standard version of the two dimensional super-Poincar\'e
transformation.\par
Section 5 extends the results of section 4 to the case of {\it mixed} fractional supersymmetry in
two dimensions. The word mixed is used to indicate that the deformation parameters of the two
braided plane coordinates on which the fractional supersymmetry is based are not necessarily at the
same root of unity. All of the algebraic and transformation properties are worked out, and as in
the supersymmetric case, spacetime Lorentz transformations are induced by suitable transformations
of the braided plane coordinates. We are thus able to introduce full two dimensional mixed
fractional super-Poincar\'e transformations. Finally we extend the
arguments of \cite{DMAP2} concerning 
the Berezin integral to the two dimensional case. 
\vfill \eject

\section{$q$-calculus for an arbitrary number of variables}
In this section we develop the $q$-calculus associated with
$r$ generalised Grassmannian variables. This calculus can
be viewed as a particular example of the braided differential
calculus described in \cite{MajI,MajII} and we present 
it from this point of view.\par
Given any matrix $\R_{12}\in \M_r\otimes \M_r$
which satisfies the quantum Yang-Baxter equation,
\be
\R_{12}\R_{13}\R_{23}=\R_{23}\R_{13}\R_{12}\q,                          \label{III.i}
\ee
as well as an associated matrix 
$\R'_{12}\in \M_r\otimes \M_r$
satisfying
\be 
\R_{12}\R_{13}\R'_{23}=\R'_{23}\R_{13}\R_{12}\q,\hskip20pt
\R_{23}\R_{13}\R'_{12}=\R'_{12}\R_{13}\R_{23}\q,                         \label{III.ii}   
\ee
\be
\R_{21}\R'_{12}=\R'_{12}\R_{21}\q,                                        \label{III.iii}
\ee
\be
({\rm P}\R+1)(P\R'-1)=0\q,                                                     \label{III.iv}
\ee
where P is the permutation matrix, we can define
a braided covector algebra \cite{MajI,MajII} with elements
$\{x_i,1\}$. This has braided Hopf algebraic 
structure given by
\be\eqalign{
\X_1\X_2&=\X_2\X_1\R'_{12}\q,\hskip10pt{\ie}\hskip10pt
x_ix_j=x_cx_b\R'\hskip2pt^{bc}\hskip0pt_{ij}\q,\cr
\Delta x_i&=x_i\otimes1+1\otimes x_i\q,\cr
\varepsilon(x_i)&=0\q,\cr
S(x_i)&=-x_i\q,\cr
\Psi_{12}(\X_1\otimes\X_2)&=\X_2\otimes\X_1\R_{12}\q,                           \label{III.v}
}\ee 
as well as $\Delta(1)=1\otimes1$, $\varepsilon(1)=1$, $S(1)=1$, 
$\Psi_{12}(1\otimes x_i)=x_i\otimes1$,
 $\Psi_{12}(x_i\otimes 1)=1\otimes x_i$.
It is convenient to use the notation $\a=\X\otimes1$, $\X=1\otimes \X$
so that the $\{w_i\}$ satisfy the same algebra as
the  $\{x_i\}$.
In this notation \cite{MajII} the coproduct $\Delta$ above is just
\be
\Delta \X=\a+\X\q,                                                           \label{III.vi} 
\ee
and the braiding $\Psi_{12}$ is equivalent to the following
braid statistics between $\X$ and $\a$
\be
\X_1\a_2=\a_2\X_1\R_{12}\q,\hskip10pt{\rm \ie}\hskip10pt
x_iw_j=w_cx_b\R\hskip2pt^{bc}\hskip0pt_{ij}\q.                            \label{III.vii}   
\ee
The notation (\ref{III.vi}) suggests that we regard
the coproduct as a generating left translations within
the braided covector space, and motivates its alternative name
{\it coaddition} \cite{MajI,MajII}.
No additional information is needed to construct the
corresponding braided left differential calculus. The 
left derivatives form a braided vector algebra, with
commutation relations give by
\be
\pl_{l1}\pl_{l2}=\R'_{12}\pl_{{\it l}2}\pl_{{\it l}1}\q,                     \label{III.viii}
\ee
and the cross relations giving their action on 
the covectors are
\be 
\pl_{l1}\X_2-\X_2\R_{21}\pl_{{\it l}1}=\delta_{12}\q.                         \label{III.ix}
\ee
The reason for the form of the second of these relations
is that along with (\ref{III.vii}) it implies that
$[\a_1\pl_{l1},\X_2]=\delta_{12}\a_1$, so that $\a_1\pl_{l1}$
can be viewed as the generator of the left translation 
(\ref{III.vi}). Equation  (\ref{III.ix}) can be viewed as giving the braiding
between the covectors and their derivatives \cite{MajIV}. To see this
we identify
\be
\Psi_{12}^{-1}(\pl_{l1}\otimes\X_2)=\X_2\otimes\R_{21}\pl_{\it{l}1}\q,                      \label{III.x}  
\ee
 so that we can rewrite (\ref{III.ix}) as
\be
\pl_{l1}\X_2-\cdot\Psi_{12}^{-1}(\pl_{l1}\otimes\X_2)=\delta_{12}\q.                    \label{III.xi}   
\ee
It is not difficult to extend this formalism
so that it includes right derivatives, and since these
play an important role in supersymmetric and 
fractional supersymmetric theories, we will do so 
explicitly. Among themselves the right derivatives 
have the same commutation relations as the left derivatives,
and thus also form a braided vector algebra so that
\be
\pl_{r1}\pl_{r2}=\R'_{12}\pl_{{\it r}2}\pl_{{\it r}1}\q.                                  \label{III.xii}
\ee
To find the cross relations between these derivatives 
and the covectors we must first reinterpret $\Delta$ as
the generator of a right shifts. 
We can do this by writing
\be
\Delta \X=\X+\ba\q,                                                           \label{III.xiii} 
\ee 
where we have introduced the alternative notation $\X=\X\otimes1$ and $\ba=1\otimes\X$.
From the braiding $\Psi_{12}$ given by (\ref{III.v}) we obtain
the braid statistics
\be
\ba_1\X_2=\X_2\ba_1\R_{12}\q.                                                   \label{III.xiv}
\ee 
In order for the right derivatives to generate the translation 
(\ref{III.xiii}) they must satisfy $[\ba_1\pl_{r1},\X_2]=\delta_{12}$.
In combination with the braid statistics (\ref{III.xiv})
this implies that
\be
\ba_1\pl_{r1}\X_2-\ba_1\X_2\R_{12}^{-1}\pl_{{\it r}1}=\delta_{12}\ba_1\q,                \label{III.xv}
\ee
from which it is clear that suitable cross relations are
\be
\pl_{r1}\X_2-\X_2\R_{12}^{-1}\pl_{{\it r}1}=\delta_{12}\q.                               \label{III.xvi} 
\ee     
As in the case of left derivatives we can interpret this as giving the
braiding between the right derivatives and the covectors.
Thus by identifying
\be \Psi_{21}(\pl_{r1}\otimes\X_2)=\X_2\otimes\R_{12}^{-1}\pl_{{\it r}1}\q,               \label{III.xvii}  
\ee
we can rewrite (\ref{III.xvi}) as
\be
\pl_{r1}\X_2-\cdot\Psi_{21}(\pl_{r1}\otimes\X_2)=\delta_{12}\q.                  \label{III.xviii}   
\ee    
Relationships (\ref{III.xi}) and (\ref{III.xviii}) motivate the 
introduction of bilinear left and right braided brackets,
\be\eqalign{
[A,B]_L&:=AB-\cdot\Psi_{12}^{-1}(A\otimes B)\q,\cr
[A,B]_R&:=AB-\cdot\Psi_{21}(A\otimes B)\q.                                       \label{III.xix} 
}\ee
The bilinearity follows from the bilinearity of $\Psi$.
This bracket is well defined on products as long as we remember
the expansion rule for the braiding. From \cite{MAJ3}
this is
\be\eqalign{
\Psi(AB\otimes C)&=(1\otimes\cdot)(\Psi\otimes1)
(A\otimes\Psi(B\otimes C))\q,\cr
\Psi(A\otimes BC)&=(\cdot\otimes1)(1\otimes\Psi)
(\Psi(A\otimes B)\otimes C)\q.                                          \label{III.xx}     
}\ee                                                          
Note that (as one would expect) $\Psi^{-1}_{12}$ and $\Psi_{21}$ expand in the same
way.
Using these brackets we can define left and right differentiation
as follows,
\be\eqalign{
\left({d\over d\X_1}\right)_L f&:=[\pl_{l1},f]_L\q,\cr
\left({d\over d\X_1}\right)_R f&:=[\pl_{r1},f]_R\q.\                   \label{III.xxi} 
}\ee
Here $f=f\{x_i\}$. We will provide a specific example shortly.
We are now able to introduce the generalised 
Grassmann algebra, which we define as the braided
 covector algebra in which $\R$ and $\R'$ are the following, $r$ dimensional matrices,
\be
\R'_{12}=\OM_{12}\q,\hskip30pt \R_{12}=\OM_{12}+(Q_1-1)\delta_{12}=T_{12}\q,        \label{III.xxii}  
\ee
the coordinate form of which is:
\be
\R'\hskip2pt^{ij}\hskip0pt_{ab}
=\om_{ab}\delta^i_a\delta^j_b\q,\hskip30pt 
\R\hskip1pt^{ij}\hskip0pt_{ab}
=(\om_{ab}+(q_a-1)\delta_{ab})\delta^i_a\delta^j_b=t_{ab}\delta^i_a\delta^j_b\q. \label{III.xxiib}  
\ee
\vskip5pt\noindent                                                     
Here $\om_{ba}=\om_{ab}^{-1}$ so that $\om_{aa}=1$,
 $\om_{ab}\neq0$, and $q_a\neq0$. It follows directly
from the fact that $\R$ and $\R'$ are diagonal 
that (\ref{III.i})-(\ref{III.iii})
are satisfied. To show that (\ref{III.iv}) 
is also satisfied we expand it explicitly
\be\eqalign{
({\rm P}\R+1)(P\R'-1)
&=\R_{21}\R'_{12}+\R'_{21}+\R_{21}-1\q,\cr
&=(\OM_{21}+(Q_2-1)\delta_{21})\OM_{12}+\OM_{21}-\OM_{21}-(Q_2-1)\delta_{21}-1\q,\cr
&=1+(Q_2-1)\delta_{21}-(Q_2-1)\delta_{21}-1\q,\cr
&=0\q.                                                                     \label{III.xxiii}        
}\ee
Putting this $\R'$ into (\ref{III.v}) we obtain the defining 
algebra of $r$ generalised Grassmann variables.
\be
\th_a\th_b=\th_j\th_i \om_{ab}\delta^i_a\delta^j_b\q,                          \label{III.xxiv}
\ee
which is equivalent to 
\be 
[\th_a,\th_b]_{\om_{ab}}=0\q.                                                  \label{III.xxv}
\ee
For left shifts we use the notation $\th_a=1\otimes\th_a$ and
$\ep_a=\th_a\otimes1$. From (\ref{III.vii}) we obtain
\be
\th_a\ep_b-\ep_j\th_i\delta^i_a\delta^j_b  t_{ij}=0\q,                       \label{III.xxvi}
\ee   
which is equivalent to 
\be  
[\ep_a,\th_b]_{t^{-1}_{ba}}=0\q.                                               \label{III.xxvii}
\ee
For the corresponding derivatives ${\cal{D}}_{La}$ we obtain 
from (\ref{III.viii}) and (\ref{III.ix}) the following commutation
and cross relations.
\be\eqalign{
[{\cal{D}}_{La},{\cal{D}}_{Lb}]_{\om_{ab}}&=0\q,\cr
[{\cal{D}}_{La},\th_b]_{t_{ba}}&=\delta_{ab}\q.                                    \label{III.xxviii}   
}\ee
For right shifts we use the notation $\th_a=\th_a\otimes1$
and $\eta_a=1\otimes\th_a$. Then from (\ref{III.xiv})  
we obtain
\be
[\eta_a,\th_b]_{t_{ab}}=0\q,                                                  \label{III.xxix}
\ee
while (\ref{III.xii}) and (\ref{III.xvi}) give us
\be\eqalign{
[{\cal{D}}_{Ra},{\cal{D}}_{Rb}]_{\om_{ab}}&=0\q,\cr
[{\cal{D}}_{Ra},\th_b]_{t_{ab}^{-1}}&=\delta_{ab}\q.                          \label{III.xxx}           
}\ee
These derivatives are generated by the left and right braided brackets, thus  from
\eq{III.xix},
\be\eqalign{
\left(\frac{d}{d\theta_a}\right)_L\theta_b&=[{\cal{D}}_{La},\theta_b]_L
=[{\cal{D}}_{La},\theta_b]_{t_{ba}}=\delta_{ab}\q,\cr
\left(\frac{d}{d\theta_a}\right)_R\theta_b&=[{\cal{D}}_{Ra},\theta_b]_R
=[{\cal{D}}_{Ra},\theta_b]_{t_{ab}^{-1}}=\delta_{ab}\q.}\ee
As an example of differentiation induced by the braided brackets 
(\ref{III.xix}) we consider the case of $r=2$, and functions $f(\theta_1,\theta_2)$
which can be expanded as positive power series of the form
\be
f(\theta_1,\theta_2)=\szit C_{l,m}\theta_1^l\theta_2^m\q.		      \label{III.xxxvi}
\ee
Then using definitions (\ref{III.xxi}) we find that
\be\eqalign{
\left({d\over d\th_1}\right)_L f(\theta_1,\theta_2)
&=[{\cal{D}}_{L1},f(\theta_1,\theta_2)]_L\q\cr
&=[{\cal{D}}_{L1},\szit C_{l,m}\theta_1^l\theta_2^m]_L\q\cr
&=\szit C_{l,m}[{\cal{D}}_{L1},\theta_1^l\theta_2^m]_{q_1^lt_{21}^m}\q\cr
&=\szit [l+1]_{q_1}C_{l+1,m}\theta_1^l\theta_2^m\q.                     \label{III.xxxvii}
}\ee
Similarly, for right differentiation we find
\be
\left({d\over d\th_i}\right)_R f(\theta_1,\theta_2)
=[{\cal{D}}_{R1},f(\theta_1,\theta_2)]_R
=\szit [l+1]_{q_1^{-1}}C_{l+1,m}\theta_1^l\theta_2^m\q.			\label{III.xxxviii}
\ee       
More generally the $C_{l,m}$ can be functions of ${\theta_j}$ where $j\neq1,2$.
This does not affect the result of the above differentiation, but of course the explicit
form given in the third line of \eq{III.xxxvii} is no longer valid. In fact, since for $a\neq b$,
$t_{ab}=t_{ba}=\omega_{ab}$, we have for $C=C(\theta_j)$ with $j\neq i$
\be\eqalign{
[{\cal{D}}_{Li},C]_L&=[{\cal{D}}_{Li},C]_{q_{ci}}=0\q,\cr
[{\cal{D}}_{Ri},C]_R&=[{\cal{D}}_{Ri},C]_{q_{ci}}=0\q,             \label{I1}
}\ee
which are the braided bracket analogues of (3.10) and (3.11). Note that unlike in the graded
bracket case considered in \cite{DMAP2} these conditions are not additional constraints on $C$, but
instead follow directly from our definition of the many variable
$q$-calculus. Note also that the conditions for induced left 
and right differentiation are compatible, so that in the many variable case, working
with braided brackets, it is not necessary to choose between these. Another difference between the graded
bracket induced derivatives of  \cite{DMAP2} and the braided bracket induced derivatives of the
present paper is that in the latter case ${\cal{D}}_{Ra}$ appears on the left of the braided
bracket. One consequence of this is that here ${\cal{D}}_{Ra}$ has a different 
normalisation. In the
many variable case, and with the new normalisation, the number and shift operators are as follows,
\be
N_a=\szi{(1-(q_a)^{m-1})\over[m]_{q_a}}\th_a^m {\cal{D}}^m_{La}
=\szi{(1-(q_a)^{1-m})\over[m]_{q_a^{-1}}}\th_a^m {\cal{D}}^m_{Ra}\q,                \label{III.xxxi}   
\ee 
\be\eqalign{
q_a^{kN_a}=\szi {1\over[m]_{q_a}}\left(\pzmm(q_a^k-q_a^l)\right)
\th_a^m {\cal{D}}_{La}^m\q,\cr
q_a^{-kN_a}=\szi {1\over[m]_{q_a^{-1}}}\left(\pzmm(q_a^{-k}-q_a^{-l})\right)
\th_a^m {\cal{D}}_{Ra}^m\q,                                                    \label{III.xxxiii}  
}\ee
\be
G_{La}=\exp_{q_a^{-1}}(\epsilon_a{\cal{D}}_{La})\q,\hskip20pt
G_{Ra}=\exp_{q_a}(\eta_a{\cal{D}}_{Ra})\q.                                       \label{I2}
\ee
These satisfy
\be
[N_a,\theta_b]=\delta_{ab}\theta_a\q,\hskip20pt
[N_a,{\cal{D}}_{Lb}]=-\delta_{ab}{\cal{D}}_{Lb}\q,\hskip20pt
[N_a,{\cal{D}}_{Rb}]=-\delta_{ab}{\cal{D}}_{Rb}\q,\hskip20pt                   \label{I3}
\ee
\be
G_{L_a}\theta_bG_{L_a}^{-1}=\delta_{ab}\epsilon_a+\theta_b\q,\hskip20pt
G_{R_a}\theta_bG_{R_a}^{-1}=\theta_b+\delta_{ab}\eta_a\q.                         \label{I4} 
\ee  
Using the identity ${\cal{D}}_{La}\theta_a-\theta_a{\cal{D}}_{La}=q_a^{N_a}$
which follows from \eq{III.xxxiii}, it is clear that with the braided bracket normalisation
the relationship between the left and right algebraic derivatives is
\be
{\cal{D}}_{Ra}=q_a^{-N_a}{\cal{D}}_{La}\q.                                    \label{III.xxxiv}
\ee
It follows immediately from this and \eq{III.xxviii} or \eq{III.xxx}  that
\be
[{\cal{D}}_{Ra},{\cal{D}}_{Lb}]_{t_{ab}}=0\q.                       \label{I20}
\ee    
Another consequence of this change of normalisation is that the $Q_a$
and $D_a$ are related by
\be
Q_a={\cal{D}}_{La}\q,\hskip20pt D_a={\cal{D}}_{Ra}\q.                    \label{I4.1} 
\ee

Left and right integrals \cite{KM} can also be introduced.
As in the one dimensional case these are defined so as to invert the effect of the corresponding 
derivatives. Another important advantage of the switch to braided brackets is that the left and
right integrals are truly distinct, and that there are simple and well defined commutation
relations amongst these as well as between them and the derivatives.
Specifically, the left integrals are defined by
\be
\int(d\th_a)_L\hskip1pt \th_a^m={\th_a^{m+1}\over[m+1]_{q_a}}\q,                  \label{III.xxxix}
\ee
and the right integrals by
\be
\int(d\th_a)_R\hskip1pt \th_a^m={\th_a^{m+1}\over[m+1]_{q_a^{-1}}}\q.             \label{III.xl}   
\ee
To integrate functions of many variables we also need the cross
relations
\be
\left[\int(d\th_a)_L,\th_b\right]_{\om_{ab}}
=\left[\int(d\th_a)_R,\th_b\right]_{\om_{ab}}=0\q,                   \label{III.xli}   
\ee
which hold for $a\neq b$.
It is also straightforward to show, for example by comparing
\be
\left(d\over d\th_a\right)_L\int(d\th_a)_R\hskip3pt\th_a^m
=\frac{[m+1]_{q_a}}{\hskip9pt[m+1]_{q_a^{-1}}}\th_a^m\q,                               \label{III.xlii}
\ee
with
\be
\int(d\th_a)_R\left(\frac{d}{d\th_a}\right)_L\th_a^m
={[m]_{q_a}\over{\hskip8pt[m]_{q_a^{-1}}}}\th_a^m\q,                            \label{III.xliii}
\ee
that the commutation relations between differentiation
and integration are as follows,        
\be
\eqalign{
\left[\left({d\over d\th_a}\right)_L,\int(d\th_b)_L\right]_{\om_{ba}}&=0\q,\hskip20pt
\left[\left({d\over d\th_a}\right)_L,\int(d\th_b)_R\right]_{t_{ba}}=0\q,\cr
\left[\left({d\over d\th_a}\right)_R,\int(d\th_b)_R\right]_{\om_{ba}}&=0\q,\hskip20pt
\left[\left({d\over d\th_a}\right)_R,\int(d\th_b)_L\right]_{t_{ab}^{-1}}=0\q.
                                                                    \label{III.xliv}
}\ee
By similar methods we also find
\def\qsss{\hskip4pt} 
\be
\left[\int(d\th_a)_L,\int(d\th_b)_L\right]_{\om_{ab}}=0\qsss,\qsss
\left[\int(d\th_a)_R,\int(d\th_b)_R\right]_{\om_{ab}}=0\qsss,\qsss
\left[\int(d\th_a)_R,\int(d\th_b)_L\right]_{t_{ab}}=0\qsss,                        \label{III.xlv}
\ee
\vskip5pt\noindent
which are the integral analogues of (\ref{III.xxviii}), (\ref{III.xxx}) 
and (\ref{I20}).\par

\section{Generalised Grassmann calculus at $q_a$ a root of unity}

One of the central results of \cite{DMAP2}, was that if
$\tilde q_a$ is a primitive $n_a$th root of unity, 
and $z_a$, $\partial_{z_a}$
are defined by 
\be
z_a=\lim_{q_a\to\tilde q_a}
{(\theta_a)^{n_a}\over[n_a]_{q_a}!}\q,\hskip20pt
\partial_{z_a}={\cal{D}}_{La}^{n_a}
=-(-1)^{n_a}{\cal{D}}_{Ra}^{n_a}\q,                            \label{IV.i}
\ee
in which \eq{III.xxxiv} has been used, and it is assumed that $(\theta_a)^{n_a}\to0$ 
as $q_a\to\tilde q_a$ in such a way that 
$z_a$ is well defined in this limit, then
\be
[\partial_{z_a},z_a]=1\q.                                      \label{IV.ii}  
\ee       
Using these definitions and the results of section 4.2
it is easy to establish the full commutation relations
in the limit as $q_a\to\tilde q_a$ (note that this limit need not be taken for all $a$).
When $q_a\to\tilde q_a$ and $q_b\to\tilde q_b$
we find from (\ref{III.xxv}), (\ref{III.xxviii}) and  (\ref{IV.ii}) 
that
\be
[\partial_{z_a},z_b]_{(t_{ba})^{n_a n_b}}
=\delta_{ab}\q,\quad
[z_a,z_b]_{(\om_{ab})^{n_a n_b}}=0\q,\quad
[\partial_{z_a}\hskip1pt,\partial_{z_b}]_{(\om_{ab})^{n_a n_b}}=0\q.  \label{IV.iii}
\ee
This clearly reduces to ordinary calculus if 
${(\om_{ba})^{n_a n_b}}=1$. It will
often be sensible to make this choice. It also follows from
(\ref{III.xxviii}), (\ref{III.xxx}) and (\ref{IV.i})
that
\be\eqalign{
[{\cal{D}}_{La},z_b]_{(t_{ba})^{n_{b}}}
&=\delta_{ab}{(\theta_{b})^{n_{b}-1}\over[n_b-1]_{q_b}!}\q,\cr
[{\cal{D}}_{Ra},z_b]_{(t_{ab})^{-n_{b}}}
&=-(-1)^{n_b}\delta_{ab}
{(\theta_{b})^{n_{b}-1}\over[n_b-1]_{q_b^{-1}}!}\q,             \label{IV.iv}  
}\ee
and that
\be
[{\cal{D}}_{La},\partial_{z_b}]_{(\om_{ab})^{n_b}}=0\q,\quad\hskip20pt
[{\cal{D}}_{Ra},\partial_{z_b}]_{(\om_{ab})^{n_b}}=0\q,\quad\hskip20pt
[\partial_{z_b},\theta_a]_{(t_{ab})^{n_b}}=0\q.              \label{IV.v}
\ee
Note that (\ref{IV.iv}) and (\ref{IV.v}) hold even when $q_a$ is not
a root of unity, as long as $q_b$ is. Following \cite{DMAP2} we can, when $q_a$ is
a root of unity, expand the algebraic {\it total} derivatives
${\cal{D}}_{La}$ and ${\cal{D}}_{Ra}$ by using the algebraic {\it partial} derivatives
$\partial_{\theta_a}$ and $\delta_{\theta_a}$. These satisfy
\vskip0pt
\be\eqalign{
&(\partial_{\theta_a})^{n_a}=(\delta_{\theta_a})^{n_a}=0\q,
\hskip20pt 
[\delta_{\theta_a},\partial_{\theta_b}]_{t_{ab}}=0\q,\cr 
&[\partial_{\theta_a},\partial_{\theta_b}]_{\om_{ab}}=0\q,\hskip48pt
[\delta_{\theta_a},\delta_{\theta_b}]_{\om_{ab}}=0\q,          \label{IV.vi}
}\ee
as well as 
\be\eqalign{
[\partial_{\theta_a},\theta_b]_{t_{ba}}&=\delta_{ab}\q,
\hskip20pt
[\partial_{\theta_a},z_{b}]_{(t_{ba})^{n_{b}}}=0\q,
\hskip20pt
[\partial_{\theta_a},\partial_{zb}]_{(t_{ba})^{-n_b}}=0\q,\cr
[\delta_{\theta_a},\theta_b ]_{(t_{ab})^{-1}}&=\delta_{ab}\q,
\hskip20pt
[\delta_{\theta_a},z_{b}]_{(t_{ab})^{-n_{b}}}=0\q,
\hskip20pt
[\delta_{\theta_a},\partial_{z_b}]_{(t_{ab})^{n_b}}=0\q.      \label{IV.vii}
}\ee
So that if we expand ${\cal{D}}_{La}$ and ${\cal{D}}_{Rb}$ as follows
\be\eqalign{
{\cal{D}}_{La}&=
\partial_{\theta_a}+
{\theta^{n_a-1}_a\over[n_a-1]_{q_a}!}\partial_{z_a}\q,\cr
{\cal{D}}_{Ra}&=
\delta_{\theta_a}-
(-1)^{n_a}{\theta^{n_a-1}_a\over[n_a-1]_{q_a^{-1}}!}\partial_{z_a}\q,   \label{IV.viii}
}\ee
then (\ref{IV.iii}) and (\ref{IV.iv}) are 
implied by (\ref{IV.vi})-(\ref{IV.viii}).\vskip0pt\noindent
If we note the identity
\be
\lim_{q_a\to{\tilde q}_a} {\theta_a^{rn_a+p}\over[rn_a+p]_{q_a}}
={z_a^{r}\hskip2pt\theta^{p}\over r!\hskip2pt[p]!}\q,                                      \label{IV.a}
\ee
then we can take the limit of (\ref{III.xxxix}) and (\ref{III.xl})
to obtain
\be
\eqalign{
\int(d\th_a)_L\hskip5pt {z_a^r\hskip2pt\th_a^p}
&=(1-\delta_{p,n-1}){z_a^r\hskip2pt\th_a^{p+1}\over[p+1]_{q_a}}
+\delta_{p,n-1}[n_a-1]_{q_a}!{z_a^{r+1}\hskip2pt\th_a^p\over (r+1)}\q,\cr
\int(d\th_a)_R\hskip5pt {z_a^r\hskip2pt\th_a^p}
&=(1-\delta_{p,n-1}){z_a^r\hskip2pt\th_a^{p+1}\over[p+1]_{q_a^{-1}}}
-(-1)^{n_a}\delta_{p,n-1}[n_a-1]_{q^{-1}_a}!{z_a^{r+1}\hskip2pt\th_a^p\over (r+1)}\q.             \label{IV.b}
}\ee
In analogy with the introduction of partial derivatives, we introduce the
following `partial' integrals,
\be
\eqalign{
\int d\th_a\hskip5pt {z_a^r\hskip2pt\th_a^p}&
=(1-\delta_{p,n-1}){z_a^r\hskip2pt\th_a^{p+1}\over [p+1]_{q_a}}\q,\cr
\int\delta\th_a\hskip5pt {z_a^r\hskip2pt\th_a^p}
&=(1-\delta_{p,n-1}){z_a^r\hskip2pt\th_a^{p+1}\over [p+1]_{q_a^{-1}}}\q,\cr
\int dz_a\hskip5pt {z_a^r}\hskip2pt\th_a^p
&={z_a^{r+1}\hskip2pt\th_a^p\over (r+1)}\q.                    \label{IV.c} 
}\ee
Using these and (\ref{IV.b}) we can write
\be
\eqalign{
\int(d\th_a)_L&=\int d\th_a+{\partial^{n_a-1}\over \partial^{n_a-1}\theta_a}\int dz_a\q,\cr
\int(d\th_a)_R&=\int \delta\th_a-(-1)^{n_a}{\delta^{n_a-1}\over \delta^{n_a-1}\theta_a}\int dz_a\q,\cr 
}\ee
which are the integral analogues of (\ref{IV.viii}). We also note the identities
\be
\int(d\th_a)_L^n=\int dz_a\q,\q 
\int(d\th_a)_R^n=-(-1)^{n_a}\int dz_a\q,\q
\int d\th_a^n=\int \delta\th_a^n=0\q.
\ee
We conclude this section with some comments on the braided Hopf structure
of the generalised Grassmann algebra and the dual algebra of derivatives as $q_a\to{\tilde q_a}$.
For an alternative derivation of \eq{IV.i} as well as a derivation of the duality properties in the
single variable case see the appendix. The results of this appendix are easily  extended to the many
variable case, and we give the results below. 
For generic $q_a$ the braided Hopf structure of $\theta_a$, 
which follows directly
from (\ref{III.v}) is as follows
\be\eqalign{
\Delta\theta_a&=\theta_a\otimes1+1\otimes\theta_a\q,\cr
\varepsilon(\theta_a)&=0\q,\cr
S(\theta_a^m)&=q^{{m(m-1)\over2}}(-\theta_a)^m\q.                 \label{IV.ix}        
}\ee
When $q_a\to{\tilde q_a}$ it follows directly from this
and (\ref{IV.i}) that in addition to (\ref{IV.ix}), which 
holds as in the
generic case,  we have the following braided Hopf structure 
for $z_a$,
\be\eqalign{
\Delta z_a&=z_a\otimes1+1\otimes z_a+
\sum_{m=1}^{n_a-1}{\th_a^m\otimes\th_a^{n_a-m}\over[m]_{q_a}![n_a-m]_{q_a}!}\q,\cr
\varepsilon(z_a)&=0\q,\cr
S(z_a)&=-z_a\q.                                                    \label{IV.x}    
}\ee
In the dual Hopf algebra with elements ${\cal D}_{L_a}$, the braided
Hopf structure is as follows
\be\eqalign{
\Delta {\cal D}_{L_a}&={\cal D}_{L_a}\otimes1+1\otimes {\cal D}_{L_a}\q,\cr
\varepsilon({\cal D}_{L_a})&=0\q,\cr
S({\cal D}_{L_a})&=q^{{m(m-1)\over2}}(-{\cal D}_{L_a})^m\q.                           \label{IV.xi} 
}\ee
The duality is given by the inner product
\be
\langle {\cal D}_{L_a},\th_b\rangle=\delta_{ab}\q,                                  \label{IV.xii}
\ee
which satisfies/is extended to products by all of the usual
identities (see appendix).
equation \eq{vseven}. 
When $q_a\to{\tilde q_a}$ the Hopf structure is extended
to include
\be
\eqalign{
\Delta\partial_{za}&=\partial_{za}\otimes1+1\otimes \partial_{za}\q,\cr
\varepsilon(\partial_{za})&=0\q,\cr
S(\partial_{za})&=-\partial_{za}\q.                                         \label{IV.xiii}
}\ee
In this case the duality is given by.
\be
\langle {\cal D}_{L_a},\th_b\rangle=\delta_{ab}\q,\q
\langle \partial_{za},z_b\rangle=\delta_{ab}\q,\q
\langle \partial_{za},\th_b\rangle=0\q,\q
\langle {\cal D}_{L_a},z_b\rangle=0\q,                                    \label{IV.xiv}       
\ee               
which follow directly from (\ref{IV.i}) and (\ref{IV.xii}).
Note that we could equally well have worked with ${\cal D}_{R_a}$, the
only advantage of using ${\cal D}_{L_a}$ being that we avoid the
factors of $(-1)^{n_a+1}$ which would arise due to \eq{IV.i}.

\section{The braided interpretation of two dimensional SUSY}

We can use the work in the previous sections to extend our new interpretation of supersymmetry
to the two dimensional case. This is of great interest in physics since it
is related to world sheet supersymmetry in superstring theory. 
The most interesting new feature in two dimensions is the presence of
Lorentz transformations. We consider a two 
dimensional generalised Grassmann algebra $\{\th_a\}$, $a=1,2$, and
its associated calculus, examining first the case of $q_1=q_2=\om_{12}=q$. 
We begin by {\it defining}
\be\eqalign{
p_\mu&=-{1\over2}{\cal{D}}_{La}(\gamma_\mu\gamma_0)_{ab}{\cal D}_{Lb}\q,\cr
x_\mu&=\lim_{q\to-1}{i\over[2]_q}\th_a(\gamma_0\gamma_\mu)_{ab}\th_b\q.             \label{V.i} 
}\ee
Here $\mu=0,1$ and $\gamma_0=\sigma_2$, $\gamma_1=i\sigma_1$, where 
${\sigma_a}$ are the usual Pauli matrices, so that we are working in the Majorana-Weyl basis for the
Dirac gamma matrices. Note that other than those 
implied by the RHS, no transformation properties are assigned to
$p_\mu$ and $x_\mu$. Since $\gamma_0\gamma_\mu$ is diagonal, the above 
can be written as 
\be\eqalign{
x_0&=i(z_1+z_2),\hskip20pt p_0=-{1\over2}(\partial_{z_1}+\partial_{z_2})\q,\cr
x_1&=i(z_1-z_2),\hskip20pt p_1={1\over2}(\partial_{z_1}-\partial_{z_2})\q,      \label{V.ii}
}\ee
from which, using (\ref{IV.iii}) it is clear that
\be
[p_\mu,x_\nu]=-ig_{\mu\nu}\q,\hskip20pt
[p_\mu,p_\nu]=0\q,\hskip20pt
[x_\mu,x_\nu]=0\q.                                           \label{V.iii}
\ee
Here $g_{\mu\nu}=diag\{1,-1\}$ so that $\{p_\mu\}$ and $\{x_\mu\}$ behave just like the quantized
momenta and coordinates of two dimensional spacetime. To establish 
their transformation properties, we proceed as follows. Under a 
translation
\be
\th_a\to\ep_a+\th_a\q,                                                    \label{V.iv}
\ee
we find from (\ref{V.i}) that the coordinates $\{x_\mu\}$ transform as follows,
\be{\eqalign{
x_\mu&\to\lim_{q\to-1}{i\over[2]_q}
(\ep_a+\th_a)(\gamma_0\gamma_\mu)_{ab}(\ep_b+\th_b)\q\cr
&=\lim_{q\to-1}{i\over[2]_q}\th_a(\gamma_0\gamma_\mu)_{ab}\th_b
+\lim_{q\to-1}{i\over[2]_q}\ep_a(\gamma_0\gamma_\mu)_{ab}\ep_b
+i\ep_a(\gamma_0\gamma_\mu)_{ab}\th_b\q\cr                       
&=x_\mu+x'_\mu+i\ep_a(\gamma_0\gamma_\mu)_{ab}\th_b\q.                      \label{V.v}
}}\ee
Together (\ref{V.iv}) and (\ref{V.v}) constitute the usual
two dimensional SUSY transformation \cite{GSW}, only now 
we can see that just as the $\{x_\mu\}$ are defined by (\ref{V.i})
in terms of the $\{\th_a\}$, the $\{x'_\mu\}$ are defined by
\be
x'_\mu=\lim_{q\to-1}{i\over[2]_q}\ep_a(\gamma_0\gamma_\mu)_{ab}\ep_b\q,        \label{V.vi}
\ee
which is the same as \eq{V.i} but with $\theta_a$ replaced by $\epsilon_a$. 
In the notation of generalised Grassmann
calculus, the infinitesimal generators of the translation 
(\ref{V.iv}) are $\epsilon_a{\cal{D}}_{La}$. On the other hand, in the usual SUSY notation,
this transformation is generated by the supercharge $Q_a$, and thus (as expected) we
can make the identification
\be
{\cal{D}}_{La}=Q_a\q.                                                           \label{V.viia}
\ee
Using this we can write the definition (\ref{V.i}) of $p_\mu$ as
\be
p_\mu=-{1\over2}Q_{a}(\gamma_\mu\gamma_0)_{ab}Q_{b}\q,                \label{V.viib}
\ee
which can easily be inverted to yield
\be
\{Q_a,Q_b\}=-2(\gamma_0\gamma_\mu)_{ab}p^\mu\q,                                  \label{V.viii}
\ee
where $p^\mu=g^{\mu\nu}p_\mu$.
Along with $[p_\mu,p_\nu]=0$ from (\ref{V.iii}), this is just
the two dimensional supersymmetry
algebra in its usual form. The usual superspace realization of
this algebra can be obtained by using (\ref{IV.viii}) and
(\ref{V.i}). We find
\be\eqalign{
Q_a={\cal{D}}_{La}&=\partial_{\th_a}+\th_a\partial_{z_a}\q\cr
&=\partial_{\th_a}-(\gamma_0\gamma_\mu)_{ab}\th_{b} p^\mu\q.                 \label{V.ix}     
}\ee
The covariant derivatives $D_a$ from two dimensional SUSY also
arise naturally in the $q\to-1$ limit of two dimensional $q$-calculus.
To see this, we write down their usual superspace realization,
\be
D_a=\partial_{\th_a}+(\gamma_0\gamma_\mu)_{ab}\th_{b} p^\mu\q.                 \label{V.x}
\ee 
Then since, using (\ref{III.xxxiii}) and (\ref{IV.vi})
\be\eqalign{
\partial_{\th_a}
&=(-1)^{N_a}\delta_{\th_a}\q\cr
&=({\cal{D}}_{La}\th_a-\th_a {\cal{D}}_{La})\delta_{\th_a}\q\cr
&=(\partial_{\th_a}\th_a-\th_a \partial_{\th_a})\delta_{\th_a}\q\cr
&=\delta_{\th_a}\q,                                                         \label{V.xi}
}\ee
we can write this as
\be
D_a=\delta_{\th_a}+(\gamma_0\gamma_\mu)_{ab}\th_{b} p^\mu\q,                   \label{V.xii}
\ee   
and thus from (\ref{IV.viii}) and (\ref{V.i}) we have
\be
D_a={\cal{D}}_{Ra}\q.         						             \label{V.xiii}
\ee
These satisfy
\be
\{D_a,D_b\}=2(\gamma_0\gamma_\mu)_{ab} p^\mu\q.                                  \label{V.IN1}
\ee
The cross relations $\{D_a,Q_b\}=0$ follow directly from \eq{I20}. 
Thus the supercharges and covariant derivatives used in two
dimensional supersymmetry, correspond respectively to
the left and right {\it total} derivatives in the $q_a\to-1$
limit of two dimensional $q$-calculus.\par
In two dimensional SUSY, the Grassmann variables ${\th_a}$ transform
as the components of a Lorentz spinor,
\be
\th_a\to S_{ab}\th_b\q,                                                         \label{V.xiv}
\ee
where
\be 
S_{ab}=\left(\matrix{\exp(\frac{\phi}{2})&0\cr                                          
0&\exp(\frac{-\phi}{2})}\right)\q.                                                  \label{V.xiva}    
\ee
 Due to 
(\ref{V.i}), the transformation properties of the coordinates $\{x_\mu\}$ are entirely
determined by those of the $\th_a$. To find these explicitly we first note that
\be
\gamma_0^2=\left(\matrix{1&0\cr0&1\cr}\right)\q,
\hskip20pt
\gamma_0\gamma_1=\left(\matrix{1&0\cr0&-1\cr}\right)\q,                           \label{V.xv}
\ee   
so that
\be\eqalign{
S\gamma_0^2 S^{T}&=\left(\matrix{\exp\phi&0\cr
0&\exp(-\phi)\cr}\right)=
\gamma_0^2\cosh{\phi}+\gamma_0\gamma_1\sinh{\phi}\q,\cr           
S\gamma_0\gamma_1 S^{T}&=\left(\matrix{\exp\phi&0\cr
0&-\exp(-{\phi})\cr}\right)=
\gamma_0^2\sinh{\phi}+\gamma_0\gamma_1\cosh{\phi}\q.                       \label{V.xvi}
}\ee
Now from (\ref{V.i}) and (\ref{V.xiv}), we find that
under a Lorentz transformation the coordinates $\{x_\mu\}$  behave as follows,
\be\eqalign{
x_\mu
&\to \lim_{q\to-1}{i\over[2]_q}\theta_b 
S_{ab}(\gamma_0\gamma_\mu)_{ac} S_{cd}\th_d\q\cr
&=\lim_{q\to-1}{i\over[2]_q}\theta_a
\Lambda_\mu\hskip0pt^\nu(\gamma_0\gamma_\nu)_{ab} \th_b\q\cr
&=\Lambda_\mu\hskip0pt^\nu x_\nu\q,                                        \label{V.xvii}
}\ee                           
in which from (\ref{V.xvi}) $\Lambda_\mu\hskip0pt^\nu$ has the form
\be
\Lambda_\mu\hskip0pt^\nu=
\left(\matrix{\cosh\phi&\sinh\phi\cr\sinh\phi&\cosh\phi\cr}\right)\q,      \label{V.xviii}
\ee 
so we have shown that, as expected, the coordinates $\{x_\mu\}$ transform like the components of a
covariant Lorentz
vector. Note that $x^\mu:=g^{\mu\nu}x_\nu$ also has the expected transformation
properties, {\it i.e.} the $\{x^\mu\}$ transform like the components of a contravariant Lorentz vector
\be
x'^\mu=(\Lambda_\nu\hskip0pt^\mu)^{-1}x^\nu\q,                                \label{V.xix}    
\ee
so that the length $x_\mu x^\mu$ is invariant. Note also that from $\{D_{La},\theta_b\}=\delta_{ab}$
and \eq{V.xiv} it follows that under a Lorentz transformation $D_{La}\to D_{Lb}S^{-1}_{ba}$, and
that through definition \eq{V.i}, this leads to the correct transformation 
properties for the $\{p_\mu\}$. 
By combining the translation and Lorentz transformation above, we can consider the effect on the coordinates 
$\{x_\mu\}$ of a general super-Poincar\'e transformation of $\theta$:
\be
\th_a\to\epsilon_a+S_{ab}\th_b\q.                                            \label{V.xx}
\ee
Under such a transformation we have, from (\ref{V.i})
\be\eqalign{
x_\mu&\to
\lim_{q\to-1}{i\over[2]_q}(\epsilon_a+S_{ab}\th_b)^2\q\cr
&=\lim_{q\to-1}{i\over[2]_q}\epsilon_a(\gamma_0\gamma_\mu)_{ab}\epsilon_b
+\lim_{q\to-1}{i\over[2]_q}\theta_b S_{ab}(\gamma_0\gamma_\mu)_{ac} S_{cd}\th_d
+i\epsilon_a(\gamma_0\gamma_\mu)_{ab}S_{bc}\th_c\q,                         \label{V.xxi}
}\ee
which by (\ref{V.v}) and (\ref{V.xvii}) reduces to
\be
x_\mu\to x'_\mu+\Lambda^\nu\hskip0pt_\mu x_\nu+
i\epsilon_a(\gamma_0\gamma_\mu)_{ab}S_{bc}\th_c\q,                              \label{V.xxii}
\ee
which is in exact agreement with the usual super-Poincar\'e transformation
of $\{x_\mu\}$.\par
Although it seems reasonable to expect that there is an analogous interpretation of 
super-Poincar\'e transformations in higher dimensions, based on \eq{V.i} or some similar
relationship, and it is indeed straightforward to construct higher dimensional algebras with
supersymmetric properties using our techniques, the generalisation of our work in this section to
$d>2$ is a nontrivial problem, and at present it remains unsolved.

\section{Mixed FSUSY in two dimensions}

Using \eq{V.ii} it is clear that in terms of $\{\th_a\}$ and $\{z_a\}$
the general 
super-Poincar\'e transformation
(\ref{V.xxii}) of the coordinates $\{x_\mu\}$ which follows from (\ref{V.xx}) takes on the following
simple form
\be
\eqalign{
z_1&\to z_1\exp\phi +z'_1+\epsilon_1\theta_1\exp(\phi/2)\q,\cr
z_2&\to  z_2\exp(-\phi)+z'_2+\epsilon_2\theta_2\exp(-\phi/2)\q.            \label{VI.i}   
}\ee 
The fact that the pairs $\{z_1,\th_1\}$ and $\{z_2,\th_2\}$ are not mixed by this 
transformation has the consequence that in this basis the generalisation
to fractional supersymmetry (FSUSY) is straightforward. To construct the most general
two dimensional FSUSY, we consider a two dimensional $q$-calculus in the limit as
$q_i\to{\tilde q}_i$, and choose $\om_{12}$ so that $\om_{12}^{n_1}=\om_{12}^{n_2}=1$. We
have included the $n_1\neq n_2$ case, and for this reason refer to our
construction as {\it mixed} FSUSY. A suitable definition for $S_{ab}$ in the 
Lorentz transformation
$\th_a\to S_{ab}\th_b$ of mixed anyonic spinors, such as $\th_a$ is
\be
S_{ab}=\left(\matrix{\exp\left(\frac{\phi}{n_1}\right)&0\cr
                     0&\exp\left(\frac{-\phi}{n_2}\right)\cr}\right)\q.                             \label{VI.ii}
\ee
As we will see, this ensures that $\{x_\mu\}$ transform as the components
of a Lorentz vector. Under a mixed anyonic Poincar\'e transformation
\be
\th_a\to\ep_a+S_{ab}\th_b\q,                                                  \label{VI.iii}
\ee
it follows from (\ref{IV.i}) that $z_1$ and $z_2$ transform as follows,
\be
\eqalign{
z_1&\to z_1\exp\phi+z_1'+\sonwm {\ep_1^m\th_1^{n_1-m}\over
[n_1-m]_{q_1}![m]_{q_1}!}\exp\left({(n_1-m)\phi\over n_1}\right)\q,\cr
z_2&\to z_2\exp(-\phi)+z_1'+\sontm {\ep_2^m\th_2^{n_2-m}\over
[n_2-m]_{q_2}![m]_{q_2}!}\exp\left({(m-n_2)\phi\over n_2}\right)\q.              \label{VI.iv}    
}\ee
To make contact with the usual spacetime coordinates $\{x_\mu\}$
we note that as in the $n_a=n_b=2$ case $z_1z_2$ is invariant under a pure Lorentz transformation
$(\ep_1=\ep_2=0)$. Thus we have $z_1z_2\propto x_0^2-x_1^2$. In fact the definitions of $x_0$ and
$x_1$ in terms of $z_1$ and $z_2$ are
\be\eqalign{
x_0&=F(z_1+z_2),\hskip20pt p_0=-{i\over2F}(\partial_{z_1}+\partial_{z_2})\q,\cr
x_1&=F(z_1-z_2),\hskip20pt p_1={i\over2F}(\partial_{z_1}-\partial_{z_2})\q,      \label{I21}
}\ee
with $F=i$ for even $n$ as in \eq{V.ii} and $F=1$ for odd $n$. These factors correspond to those
relating $t$ to $z$ in \cite{DMAP1,DMAP2,DMAP3,DMAP4} and ensure the
reality of $p_\mu$ and $x_\mu$. From  (\ref{IV.iii})
it follows that the operators defined by \eq{I21} satisfy (\ref{V.iii}) as in the supersymmetric
case covered in the last section. After a
little algebra we obtain the mixed anyonic transformation of the $\{x_\mu\}$ coordinates
\be
x_\mu\to x'_\mu+\Lambda^\nu\hskip0pt_\mu x_\nu+\sum_{a,b=1}^2\sonam 
{F\epsilon_a^m(\gamma_0\gamma_\mu)_{ab}(S_{bc}\th_c)^{n_a-m}\over
[n_a-m]_{q_a}![m]_{q_a}!}\q.                                                 \label{VI.v}
\ee
Here $\Lambda^\nu\hskip0pt_\mu$ is the same as in (\ref{V.xviii}).
The fractional supercharge and covariant derivative are also easy to
deduce. From (\ref{IV.viii}) and (\ref{I21}) we find
\be
\eqalign{
Q_a&={\cal{D}}_{La}=\partial_{\theta_a}+
{\theta^{n_a-1}_a\over[n_a-1]_{q_a}!}\partial_{za}\q\cr
&=\partial_{\theta_a}+{iF\over [n_b-1]_{q_b}!}
(\gamma_0\gamma_\mu)_{ab}\th_b^{n_b-1}p^\mu\q,                            \label{VI.vi}  
}\ee
and
\be
\eqalign{
D_a&={\cal{D}}_{Ra}=\delta_{\theta_a}-
(-1)^{n_a}{\theta^{n_a-1}_a\over[n_a-1]_{q_a^{-1}}!}\partial_{za}\q\cr
&=\delta_{\theta_a}-{iF(-1)^{n_a}\over [n_b-1]_{q^{-1}_b}!}
(\gamma_0\gamma_\mu)_{ab}\th_b^{n_b-1}p^\mu\q.                       \label{VI.vii}
}\ee
The algebraic (left) integral of a function $f(z_1,z_2,\theta_1,\theta_2)$ 
on 2-dimensional fractional
superspace is
\be
\int(d\theta_2)_L\int(d\theta_1)_L\hskip5ptf(z_1,z_2,\theta_1,\theta_2)\q.    \label{VI.viii} 
\ee
Note that this integral would change by an overall multiplicative factor 
if we reversed the order of $\int(d\theta_2)_L$ and $\int(d\theta_1)_L$, 
so that in writing down (\ref{VI.viii}), we have made a choice of convention.
This integral can be expanded using \eq{IV.b} to yield
\be
\int d\theta_2\int d\theta_1\hskip3pt f+\int d\theta_2
{\partial^{n_1-1}\over\partial^{n_1-1}\theta_1}\int dz_1\hskip3pt f
+{\partial^{n_2-1}\over \partial^{n_2-1}\theta_2}\int dz_2\int d\theta_1\hskip3pt f
+{\partial^{n_2-1}\over
\partial^{n_2-1}\theta_2}{\partial^{n_1-1}\over\partial^{n_1-1}\theta_1}\int dz_2\int dz_1\hskip3pt f
\label{I10}
\ee                
\vskip5pt\noindent
To obtain a numerical measure from this algebraic integral
we now make use of an argument similar to that given in
\cite{DMAP2}. $\theta_1$ and $\theta_2$ are nilpotent and thus all of
their eigenvalues are zero. On the other hand, the bosonic limits
denoted by $z_1$ and $z_2$ are non-nilpotent and thus do have
non-zero eigenvalues. After integration, the first three terms in \eq{I10}
always involve $\theta_1$ or $\theta_2$ raised to some non-zero power,
whereas the last term involves $z_1$ and $z_2$ only. Any numerical
measure based on the integral \eq{I10} must be based on its
eigenvalues in some representation. Consequently only the last
term contributes and thus the first three can be dropped.
 It is convenient at this point to introduce 
a fractional Berezin integral as in \cite{DMAP2}.
\be
\int\hskip5pt (d\theta_a)_{Ber}
={\partial^{n_a-1}\over\partial^{n_a-1}\theta_a}\q,                           \label{VI.ix} 
\ee  
The resulting numerical integral measure on two dimensional fractional
superspace can now be written as
\be
I(f)=\int dz_2dz_1(d\theta_2)_{Ber}(d\theta_1)_{Ber}
\hskip5pt f(z_1,z_2,\theta_1,\theta_2)\q.                                   \label{VI.x}    
\ee
If we expand $f$ as a power series
\be
f(z_1,z_2,\theta_1,\theta_2)
=\sum_{m_1=0}^{n_1-1}\sum_{m_2=0}^{n_2-1}
C_{m_1,m_2}(z_1,z_2){\theta_1^{m_1}
\over[m_1]_{q_1}!}{\theta_2^{m_2}\over[m_2]_{q_2}!}\q,                                \label{VI.xi}
\ee 
then (\ref{VI.x}) reduces to
\be
I(f)=\int dz_2dz_1\hskip3pt C_{n_1-1,n_2-1}(z_1,z_2)\q.                  \label{VI.xii}   
\ee
Note that up to a constant Jacobian factor this is equal to 
\be
\int dx_0dx_1\hskip3pt C_{n_1-1,n_2-1}(z_1,z_2)\q,                   \label{VI.xiii}     
\ee
which, for $n$=2, is just the integral which arises in supersymmetric 
field theories involving one space and one time dimension.

\def\be{\begin{equation}}
\def\ee{\end{equation}}
\def\Comm#1{{\tt{[#1]}}}
\def\sp{\hskip2pt}
\def\li{q\to-1}
\def\ff{f(\theta)}
\def\drr{{d\over d\theta}}
\def\ep {\parindent=0pt\vfill\eject}
\def\np {\parindent=20pt}
\def\s {\vskip0pt\noindent}
\def\bb #1 {\vskip0pt\noindent[#1]\hskip5pt}
\def\r #1 {\hskip5pt[#1]\hskip3pt}
\def\exp{{\rm exp}}
\def\ii{{\rm id}}
\def\a{{a^\dagger}}
\def\b{{b^\dagger}}
\def\tta{{\theta^{n-1}\over[n-1]_q!}}
\def\sn{{\sum_{m=0}^{n-1}}}
\def\sr{{\sum_{m=0}^{r}}}
\def\sno{{\sum_{m=0}^{n-1}}}
\def\pt{{\partial\over\partial\theta}}
\def\dt{{\partial\over \partial t}}
\def\dr{{d_R\over d_R\theta}}
\def\dl{{d_L\over d_L\theta}}
\def\dz{\partial_t}
\def\qq{\exp({2\pi i\over n})}
\def\qm{\lim_{q\to-1}}
\def\ql{\lim_{q\to\qq}}
\def\qmm{{\lim\over{q\to-1}}}
\def\ad{a^\dagger}
\def\ff{f(\theta)}
\def\zz{\vert0\rangle}
\def\rr{{d_R\over d_R\theta}}
\def\ll{{d_L\over d_L \theta}}
\def\gg{{q^{g(A_m)}}}
\def\ssn{{\sum_{n=0}^\infty}}
\def\ss{{\sum_{m=0}^\infty}}
\def\d{{{\cal D}_L}}
\def\dR{{{\cal D}_R}}
\def\ie{{\it i.e.}}
\def\dual{{\cal K}}

\appendix
\begin{Large}
\vskip20pt\noindent
{\bf Appendix}
\end{Large}
\vskip10pt
\smal
\setlength{\abovedisplayskip}{13pt}
\setlength{\belowdisplayskip}{13pt}
\setlength{\belowdisplayshortskip}{13pt}
\jot10pt
\sss
\def\der{{\cal D}_L}
\def\q{\quad}
The results of \cite{DMAP1,DMAP2,DMAP3,DMAP4} can also be derived from a different and in some ways
mathematically nicer point of view. Our work here uses a technique
similar to that employed by G.I. Lusztig in his work on the properties
of deformed universal enveloping algebras with deformation parameter equal to a
root of unity \cite{LUSZTIG1,LUSZTIG2}. To the best of our knowledge this is the first time
such a technique has been applied to a braided object.
 Let us begin by introducing the braided Hopf
algebra ${\cal A}$, which we define for all $q$. This has elements $\{\theta^{(m)}\}$,
$m=0,1,2,\ldots,\infty$ with $\theta^{(0)}=1$, and relations
\be
\theta^{(m)}\theta^{(p)}=\frac{[m+p]_q!}{[m]_q![p]_q!}\theta^{(m+p)}\q,         \label{vone}
\ee
as well as
\be\eqalign{
\Delta\theta^{(m)}&=\sum_{r=0}^m\theta^{(m-r)}\otimes\theta^{(r)}\q,\cr
\varepsilon(\theta^{(m)})&=\delta_{m,0}\q,\cr
S(\theta^{(m)})&=(-1)^mq^{\frac{m(m-1)}{2}}\theta^{(m)}\q.               \label{vtwo}
}\ee
The braiding is given by
\be
\psi(\theta^{(m)}\otimes\theta^{(s)})=q^{ms}
\theta^{(m)}\otimes\theta^{(s)},
\label{vtwopfive}
\ee
so that
\be
(\theta^{(r)}\otimes\theta^{(m)})(\theta^{(s)}\otimes\theta^{(s)})=
q^{ms}\theta^{(r)}\theta^{(s)}\otimes\theta^{(m)}\theta^{(t)}\q.              \label{vthree}
\ee  
We also define $\dual$ the braided Hopf algebra dual to ${\cal A}$ as follows.
This has elements $\{\der^{(m)}\}$,
$m=0,1,2,\ldots,\infty$ with $\der^{(0)}=1$, and relations
\be
\der^{(m)}\der^{(p)}=\der^{(m+p)}\q,                                       \label{vfour}
\ee
as well as
\be\eqalign{
\Delta\der^{(m)}&=\sum_{r=0}^m\frac{[m]_q!}{[r]_q![m-r]_q!}\der^{(m-r)}\otimes\der^{(r)\q,}\cr
\varepsilon(\der^{(m)})&=\delta_{m,0}\q,\cr
S(\der^{(m)})&=(-1)^mq^{\frac{m(m-1)}{2}}\der^{(m)}\q.               \label{vfive}
}\ee
The braiding is given by
 \be
\psi(\der^{(m)}\otimes\der^{(s)})=q^{ms}
\der^{(m)}\otimes\der^{(s)},
\label{vsix}
\ee 
These two braided Hopf algebras are dual in the sense that there is a bilinear map 
$\langle\hskip3pt,\hskip2pt\rangle:{\cal A}\otimes\dual\mapsto$ the complex plane, such that
\be\eqalign{
\langle a,xy\rangle&=\langle\Delta a,x\otimes y\rangle\q,\cr
\langle ab,x\rangle&=\langle a\otimes b,\Delta x\rangle\q,\cr
\langle 1,x\rangle&=\varepsilon_\dual(x)\q,\cr
\langle a,1\rangle&=\varepsilon_{\cal A}(a)\q,\cr
\langle S(a),x\rangle&=\langle a,S(x)\rangle\q.                        \label{vseven}
}\ee
Specifically, in this case we have
\be
\langle\theta^{(m)},\der^{(p)}\rangle=\delta_{mp}\q,                  \label{veight}
\ee
the compatibility of which with (\ref{vseven}) is easy to verify.
We now consider the cases of generic $q$ and $q$ a root of unity separately.\vskip0pt\noindent
i) generic $q$ or $q=1$. From (\ref{vone}) it follows that
\be\eqalign{
\theta^{(m)}&=\frac{\theta^{(1)}\theta^{(m-1)}}{[m]_q}=
\frac{\theta^{(1)2}\theta^{(m-2)}}{[m]_q[m-1]_q}\q\cr
&=\frac{(\theta^{(1)})^m}{[m]_q!}\q,                                           \label{vnine}
}\ee
and similarly
\be
\der^{(m)}=(\der^{(1)})^m\q.                                                      \label{vten} 
\ee
Consequently, at generic $q$ the braided Hopf algebra ${\cal A}$ and its dual $\dual$ 
are both finite dimensional, each containing the identity, and only one other element. If we define
$\theta=\theta^{(1)}$ and $\der=\der^{(1)}$ then we can write the generic $q$ braided Hopf 
structure as follows. For ${\cal A}$ we have
\be\eqalign{
\Delta\theta&=\theta\otimes1+1\otimes\theta\q,\cr
\varepsilon(\theta)&=0\q,\cr
S(\theta)&=-\theta\q,                                                    \label{veleven}
}\ee
which recovers the braided line at generic $q$, and for $\dual$ we have
\be\eqalign{
\Delta\der&=\der\otimes1+1\otimes\der\q,\cr
\varepsilon(\der)&=0\q,\cr
S(\der)&=-\der\q.                                                        \label{vtwelve}
}\ee   
The duality simplifies to
\be
\langle\theta,\der\rangle=1\q,                                               \label{vthirteen}
\ee
and is extended to products via (\ref{vseven}).
By comparing (\ref{veleven}) and (\ref{vtwelve}) with one of  \cite{MajI,MajII,MajIII} we are able to
identify both the braided hopf algebra $\cal A$ and its dual $\dual$ with
the braided line when $q$ is not a root of unity. 

\vskip0pt\noindent
ii) $q$ a primitive $n$th root of unity. As in the generic $q$ case we can use 
(\ref{vone}) to obtain
\be
\theta^{(p)}=\frac{(\theta^{(1)})^p}{[p]_q!}\q,                            \label{vfourteen}
\ee
but since $[n]_q=0$ this only works for $p=0,1,\ldots,n-1$. 
However we are able to write
\be\eqalign{
\theta^{(rn+p)}&=\theta^{(rn)}\theta^{(p)}\lim_{q\to\epsilon}
\frac{[rn]_q![p]_q!}{[rn+p]_q!}\q\cr
&=\theta^{(rn)}\theta^{(p)}\q,                                         \label{vfifteen} 
}\ee
where $r\geq0$ and $0\leq p\leq n-1$. Also using (\ref{vone})
we find that
\be\eqalign{
\theta^{(rn)}&=\theta^{(n)}\theta^{((r-1)n)}\lim_{q\to\epsilon}
\frac{[(r-1)n]_q![n]_q!}{[rn]_q!}\q\cr
&=\frac{\theta^{(n)}\theta^{((r-1)n)}}{r}\q.                        \label{vsixteen}
}\ee
Iterating we finally obtain
\be
\theta^{(rn)}=\frac{{(\theta^{(n)})}^r}{r!}\q,                         \label{vseventeen}
\ee
so that
\be
\theta^{(rn+p)}=\frac{{(\theta^{(n)})}^r\theta^{(p)}}{r!}\q.           \label{veightteen}  
\ee 
Similarly, for the dual we find that
\be
\der^{(rn+p)}=(\der^{(n)})^r\der^{(p)}\q.                                 \label{vnineteen}  
\ee   
Thus when $q$ is a root of unity ($q\neq1$) the braided Hopf algebra 
${\cal A}$ is finite dimensional, having two independent elements 
$\theta^{(1)}$ and $\theta^{(n)}$ besides the identity. The dual $\dual$
is also finite dimensional, but it has only one independent element $\der^{(1)}$
besides the identity. It is convenient to define 
\be
\theta=\theta^{(1)}\q,\hskip10pt 
z=\theta^{(n)}\q,\hskip10pt
\der=\der^{(1)}\q,\hskip10pt 
\partial_z=\der^{(n)}\q.                                               \label{vtwenty} 
\ee
Using this notation, the algebraic relations (\ref{vone}) reduce to 
$[\theta,z]=0$ and $\theta^n=0$,
and the braided Hopf structure (\ref{vtwo}) reduces to
\be
\eqalign{
\Delta\theta&=\theta\otimes1+1\otimes\theta\q,\cr
\varepsilon(\theta)&=0\q,\cr
S(\theta)&=-\theta\q,\cr                                                    \label{vtwentyone}
}\ee
and
\be\eqalign{
\Delta z&=z\otimes1+1\otimes z+\sum_{m=1}^{n-1}\frac{\theta^m\otimes\theta^{n-m}}
{[n-m]_q![m]_q!}\q,\cr
\varepsilon(z)&=0\q,\cr
S(z)&=-z\q.                                           \label{vtwentytwo} 
}\ee  
The braided Hopf structure of the dual $\dual$ is given by
\be
\eqalign{
\Delta\der&=\der\otimes1+1\otimes\der\q,\cr
\varepsilon(\der)&=0\q,\cr
S(\der)&=-\der\q.                                                        \label{vtwentythree}
}\ee   
which, using $\partial_z=\der^n$ implies the following braided
Hopf structure for $\partial_z$,
\be\eqalign{
\Delta\partial_z&=\partial_z\otimes1+1\otimes\partial_z\q,\cr
\varepsilon(\partial_z)&=0\q,\cr
S(\partial_z)&=-\partial_z\q.                                           \label{vtwentyfour} 
}\ee  
The duality (\ref{veight}) now takes on the form
\be
\langle z^r\theta^p,\der^{r'n+p'}\rangle=\langle z^r\theta^p,\partial_z^r\der^{p'}\rangle
=\delta_{r,r'}\delta_{p,p'}\hskip1ptr!\hskip1pt[p]_q!\q,                 \label{vtwentyfive}
\ee 
so that in particular
\be
\langle\theta,\der\rangle=1\q,\hskip20pt\langle z,\partial_z\rangle=1\q.       \label{vtwentysix}   
\ee
Thus when $q$ is a root of unity ${\cal A}$ coincides with the braided Hopf algebra
which was associated in previous work with a limit of the braided line
as its deformation parameter goes to a root of unity. Using this approach we have also
obtained the braided Hopf structure of the dual and details of the duality when $q$ is a 
root of unity (this is an alternative form of the braided line when
$q$ is a root of unity). Note also that the $\theta$ part of $\cal A$
forms a braided sub-Hopf algebra, but that the $z$ part does not.\par
The advantage of the approach adopted here is that it  enables us to
restrict the taking of limits to purely numerical
quantities, for which they are manifestly well defined. In this appendix we have 
worked with left derivatives ${\cal{D}}_L$ only, but we could equally well have 
chosen right derivatives ${\cal{D}}_R$, for which a closely analogous
treatment exists.\par 
The relationship between the work in this appendix and the work of
\cite{LUSZTIG1,LUSZTIG2} suggests that the latter might also have a
physical interpretation in terms of supersymmetry and fractional
supersymmetry. This idea will be developed further in \cite{RSD}. 
\vskip30pt\noindent
\begin{Large}
{\bf Acknowledgements}
\end{Large}
\vskip10pt

{ Thanks to Alan Macfarlane, Jos\'e de Azc\'arraga and Carlos P\'erez
Bueno for many helpful discussions and communications. I gratefully
acknowledge the financial support provided by the E.P.S.R.C. and  St
John's College, Cambridge during the preparation
of this paper.}

\end{document}